\documentclass[a4paper]{article}

\usepackage[top=2.8cm,bottom=3cm,left=3cm,right=3cm]{geometry}
\usepackage{graphicx}
\usepackage{txfonts}
\usepackage{amsfonts}
\usepackage{amssymb}
\usepackage{booktabs}
\usepackage{xcolor}
\usepackage{lineno}
\usepackage{hyperref}

\usepackage{natbib}

\begin{document}


\Large
\textbf{\begin{center}{CLEAN and multi-scale CLEAN for STIX in Solar Orbiter} \end{center}}

\normalsize

Miriana Catalano$^{1}$, Anna Volpara$^{1}$, Paolo Massa$^{2}$, Michele Piana$^{1,3}$ and Anna Maria Massone$^{1,3}$ \\

\hspace{-0.5cm}$^1$ MIDA, Dipartimento di Matematica, Universit\`a di Genova, via Dodecaneso 35 16146 Genova, Italy  \\
$^2$ Institute for Data Science, University of Applied Sciences and Arts Northwestern Switzerland, Bahnhofstrasse 6, Windisch, 5210, Switzerland \\
$^3$ Istituto Nazionale di Astrofisica, Osservatorio Astrofisico di Torino, via Osservatorio 20 10025 Pino Torinese Italy \\


\begin{center}
\textbf{Abstract}
\end{center}
CLEAN is a well-established deconvolution approach to Fourier imaging at both radio wavelwengths and hard X-ray energies. However, specifically for hard X-ray imaging, CLEAN suffers two significant drawbacks: a rather limited degree of automation, and a tendency to under-resolution. This paper introduces a multi-scale version of CLEAN specifically tailored to the reconstruction of images from measurements observed by the Spectrometer/Telescope for Imaging X-rays (STIX) on-board Solar Orbiter. Using synthetic STIX data, this study shows that multi-scale CLEAN may represent a reliable solution to the two previously mentioned CLEAN limitations. Further, this paper shows the performances of CLEAN and its multi-scale release in reconstructing experimental real scenarios characterized by complex emission morphologies.
 \\

\textbf{keywords:} solar flares; hard X-ray solar imaging; Fourier techniques


\section{Introduction} \label{sec:intro}
The imaging concept for the most recent space missions in solar hard X-ray imaging is a Fourier transform model: native measurements are samples of the Fourier transform of the incoming photon radiation, named visibilities, and image reconstruction in this framework consists therefore in solving the Fourier transform inversion problem from limited data \citep{bertero2006regularization}. Past examples of instruments that rely on this imaging approach are the Hard X-ray Space Telescope (HST) on-board YohKoh \citep{kosugi1992hard}, which utilized 32 pairs of modulation collimators to measure 32 visibilities in four broad bands in the energy range between 14 and 100 keV; and the Reuven Ramaty High Energy Solar Spectroscopic Imager (RHESSI) \citep{2004SPIE.5171...38L}, which observed hard X-rays and gamma-rays by means of nine rotating modulation collimators generating hundreds of visibilities with heterogeneous signal-to-noise ratio. Two more Fourier-based space telescopes are currently flying to record hard X-ray observations of the Sun, both characterized by a hardware technology based on Moiré patterns. Specifically, the Spectrometer/Telescope for Imaging X-rays (STIX) on-board Solar Orbiter \citep{2012SPIE.8443E..3LB,2020A&A...642A..15K}, and the Hard X-ray Imager (HXI) on-board ASO-S \citep{2019RAA....19..160Z} are currently measuring 30 and 45 visibilities, respectively, although from two completely different orbits.

Within the framework of the RHESSI mission, several imaging methods have been developed, all designed to extract spatial information from the measured either counts or visibilities \citep{2002SoPh..210...61H,2009ApJ...698.2131D,2007SoPh..240..241S,2013A&A...555A..61B,2006ApJ...636.1159B,2020ApJ...894...46M,2009ApJ...703.2004M,2017ApJ...849...10F,2018A&A...615A..59D,2008ApJ...677..704H,2002SoPh..210..193A,1996ApJ...466..585M}. In particular, image deconvolution based on CLEAN \citep{1974A&AS...15..417H,2002SoPh..210...61H} was the imaging tool applied for more than 10$\%$ of RHESSI studies, which made this technique the most popular one in the RHESSI community. The steps of this algorithm are made of a CLEAN loop generating a set of CLEAN components located where most of the source emission propagates; an estimate of the background; the sum of the CLEAN component map with the estimated background; and, eventually, the generation of the
CLEANed map via convolution with an idealized Point Spread Function (PSF) named the CLEAN beam.

However, it is well-established that the CLEAN algorithm developed for RHESSI was affected by two main weaknesses:
\begin{itemize}
    \item The degree of automation of CLEAN was rather limited, the final convolution step being significantly dependent on the user's choice.
    \item The maps reconstructed by CLEAN are typically under-resolved, with a rather poor ability to identify different spatial scales in the field of view under analysis.
\end{itemize}
On the one hand, the first limitation has been addressed in \cite{2023ApJS..268...68P}, which introduced an unbiased, user-independent version of CLEAN for STIX. On the other hand, the formalism of a possible multi-scale release of CLEAN has been outlined in \cite{2024InvPr..40l5017V} for the solution of the general Fourier transform inversion problem from limited data. That paper also briefly illustrates an application to RHESSI observations. 

The objective of the present paper is to introduce the CLEAN and multi-scale CLEAN algorithms into the STIX framework, and to validate their performances against both synthetic data and experimental observations. Specifically, for this latter application, we considered the flaring events occurred on February, 24 2024, March, 10 2024, and May 14 2024, and compared the reconstructions provided by both CLEAN and multi-scale CLEAN in the case of STIX data, with images of the same events provided by the Atmospheric Imaging Assembly (AIA) on-board the Solar Dynamics Observatory in the EUV regime \citep{2012SoPh..275...17L} and by the Hard X-ray Imager (HXI) on-board the Advanced Space-Based Solar Observatory (ASO-S), again in the case of hard X-ray energies \citep{2019RAA....19..160Z}. 

The plan of the paper is as follows. Section 2 illustrates how CLEAN and multi-scale CLEAN can be designed to deal with STIX data. Section 3 validates the multi-scale approach in the case of synthetic visibilities. Section 4 applies multi-scale CLEAN to the analysis of STIX measurements recorded during three flaring events in 2024. Our conclusions are offered in Section 5.

\section{CLEAN in the STIX framework}

CLEAN is a deconvolution algorithm that has been introduced by \cite{1974A&AS...15..417H} in the case of radioastronomy imaging and that has been translated into a solar hard X-ray context (\cite{2002SoPh..210...61H}) for the processing of data recorded by the Reuven Ramaty High Energy Solar Spectroscopic Imager (RHESSI). The basic idea behind CLEAN is that the image to be reconstructed can be represented as the superposition of point sources that are iteratively introduced in the reconstruction process. More schematically, CLEAN is implemented by means of the following steps:
\begin{enumerate}
    \item A dirty map of the hard X-ray emission is obtained by applying the Discrete Fourier Transform to the set of visibilities recorded during a specific time interval and in correspondence with a specific photon energy channel.
    \item After identifying the maximum in the dirty map, a CLEAN component map is generated by means of a $\delta$-Dirac placed at the maximum position.
    \item The dirty beam, i.e. the instrument Point Spread Function (PSF), centered at the maximum position is then subtracted from the dirty map.
\end{enumerate}
Items 2 and 3 are iterated until the remaining dirty map contains just noise, and the final CLEANed image is obtained by adding to the CLEAN component map an estimate of the residuals and then convolving with the so-called CLEAN beam, i.e., an idealized form of the instrument PSF obtained by fitting the dirty beam by means of an idealized two-dimensional Gaussian function. 

\begin{figure*}
    \centering   \includegraphics[width=0.98\linewidth]{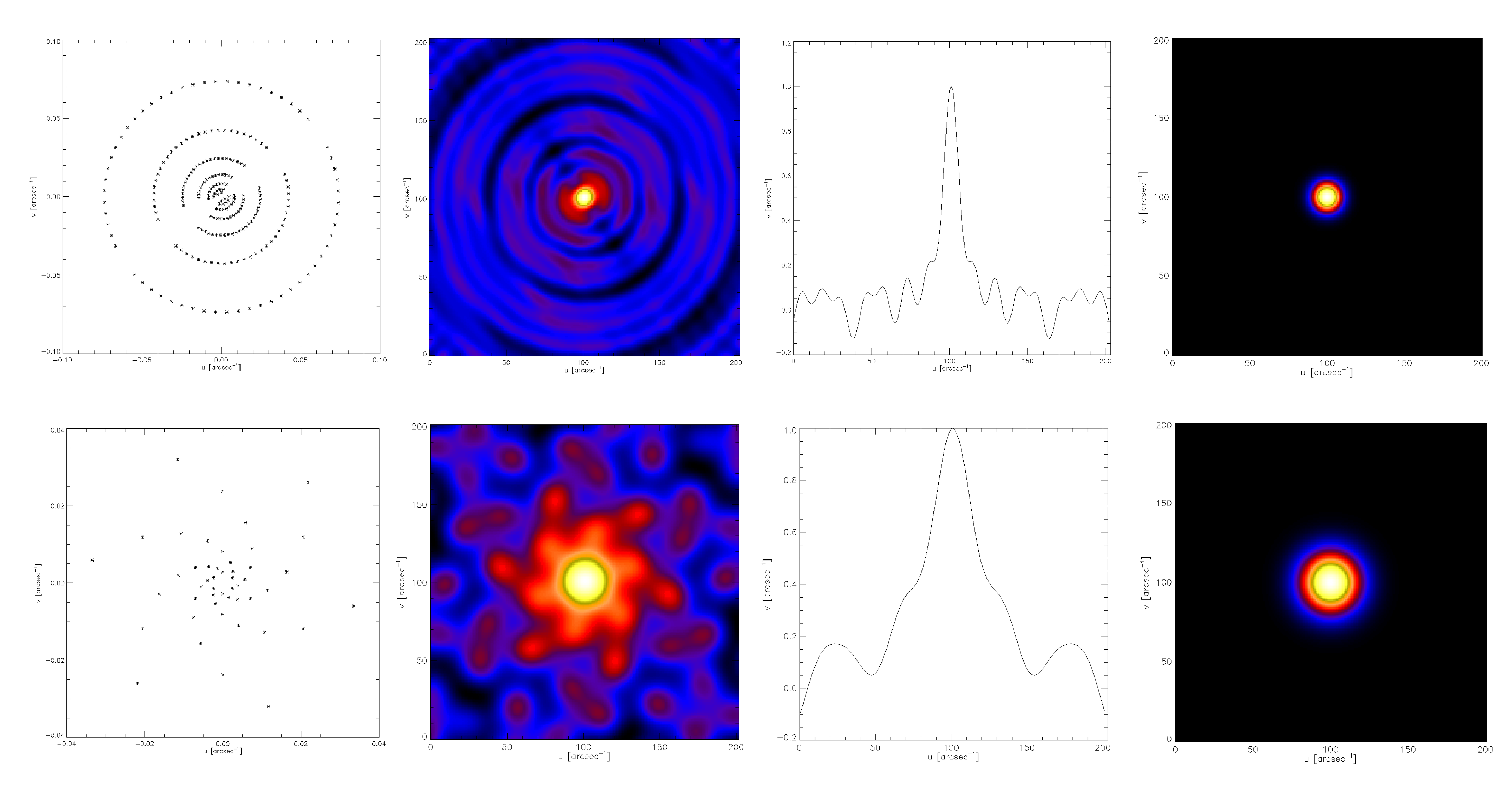}
    \caption{Comparison between the imaging characteristics of RHESSI (top row) and STIX (bottom row). First column: sampling of the $(u,v)$ plane; second column: Point Spread Functions (PSFs); third column: central cut of the PSFs; fourth column: idealized PSFs obtained by fitting the central peak of the original PSFs.}
    \label{fig:RHESSI-STIX}
\end{figure*}

As shown in Figure \ref{fig:RHESSI-STIX}, this scheme depends on the properties of the telescope at different levels, which impacts its implementation within the STIX framework. First, the sampling of the spatial frequency plane (the $(u,v)$ plane hereafter) is significantly different; this implies different properties of the instrumental PSFs and, specifically, of the shapes of their central peaks, which are a measure of the spatial resolution achievable by the two instruments. As a consequence, this difference in the resolution power is reflected by the different Full Width at Half Maximum (FWHM) of the RHESSI and STIX CLEAN beam. Specifically, on the one hand the nominal maximum RHESSI angular resolution was $2.26$ arcsec. On the other hand, STIX achieves a nominal maximum resolution of $7.2$ arcsec, although at the current stage of the calibration process collimators 1 and 2 are not included in the imaging process, so that the maximum resolution decreases to $14.6$ arcsec. However, Solar Orbiter can approach the Sun as close as 0.28 astronomical units (AU). At this distance, an angular resolution of 1 arcsecond corresponds to a linear scale of about 200 km on the solar surface, compared to roughly 725 km per arcsecond at 1 AU (the Earth–Sun distance, relevant for RHESSI). Thus, although the nominal angular resolution of STIX is lower than that of RHESSI, the much closer observing distance effectively compensates for this limitation, providing a comparable — or even superior — spatial resolving power on the Sun.

The experience in using CLEAN for the analysis of RHESSI visibilities showed that this imaging approach has two main limitations. First, CLEAN is not capable of automatedly adapting to the spatial scale of the source to be reconstructed, which often results in under-resolution effects in reconstructions. Second, the step implementing the convolution between the CLEAN component map and the CLEAN beam is significantly user's dependent, since the FHWM of the CLEAN beam is typically tuned according to heuristic considerations. Very recently, \cite{2024InvPr..40l5017V} formulated a multi-scale version of CLEAN, which does not require the final convolution step and which was validated against both synthetic and experimental RHESSI data. As for RHESSI, also in the case of STIX this multi-scale CLEAN method explicitly leverages the fact that the sampling of the Fourier transform of the incoming flux made by the ESA telescope is made over circles in the $(u,v)$ plane. 

Specifically, the starting points for multi-scale CLEAN in the case of STIX visibilities are, first, to group the set of $10$ circles characterizing the STIX $(u,v)$ plane into $N$ subsets, and to model the image $I(x,y)$ to reconstruct according to the formula
\begin{equation}\label{eq:model-equation}
I(x,y) = \sum_{i=1}^N\sum_{q_i=1}^{Q_i}I_{q_i}m_i(x-x_{q_i},y-y_{q_i}) + B(x,y) \ \ ,
\end{equation}
which assumes that, for each subset, there exists an unknown image that can be modelled as the sum of $Q_i$ normalized basis functions $m_i(x,y)$, each one centered at $(x_{q_i},y_{q_i})$ and with peak intensity $I_{q_i}$. Second, for each subset, the corresponding dirty map is computed by applying the inverse Fourier transform to the Fourier samples corresponding to each subset. Third, again for each subset, the corresponding PSF is computed. If we denote by $CC^{(0)}(x,y)$ a map made of pixels with zero content, the first iteration is computed according to a process that reads as follows:
\begin{enumerate}
    \item Each dirty map $I^{(0)}_j$, $j=1,\ldots,N$, is rescaled according to the rule described in \cite{2024InvPr..40l5017V} .
    \item The first selected scale is identified among all dirty maps, as the one corresponding to the re-scaled dirty map presenting the maximum intensity value (note that re-scaled dirty maps are used just in this step, i.e., for maximum identification). Then, the corresponding PSF $K(x,y)$ and the corresponding basis function $m(x,y)$ are selected.
\item The multi-scale CLEAN components map is computed as
\begin{equation}\label{multi-step-2}
CC (x,y) = CC^{(0)}(x,y) + \frac{ \gamma I_{max}}  {\max\limits_{ (x,y) }   |(m * K)(x,y)|}m(x-x_{max},y-y_{max}),
\end{equation}
where $I_{max}$ is the maximum value of the dirty map at the selected scale, $(x_{max},y_{max})$ denotes its position, and $\gamma$ is a {\em{gain factor}}.
\item At each scale $j=1,\ldots,N$, the dirty map component is computed in such a way that
\begin{equation}\label{multi-step-3}
I_j(x,y) = I^{(0)}_j(x,y) - \frac{\gamma I_{max}}{\max\limits_{(x,y)} |(m * K)(x,y)|} (m*K_j)(x-x_{max},y-y_{max}).
\end{equation} 
\end{enumerate}
Iterations continue while accordingly updating $CC(x,y)$ until a stopping rule is satisfied. 

In the final step of multi-scale CLEAN, no convolution with any CLEAN beam is needed, and one simply constructs the CLEANed map $C(x,y)$ as 
\begin{equation}\label{multiCLEAN-final}
C(x,y) = CC^{(L)}(x,y)+B^{\prime}(x,y) \ \ ,
\end{equation}
where $CC^{(L)}(x,y)$ is given by (\ref{multi-step-2}) at the last iteration and $B^{\prime}(x,y)$ is an estimate of the background.

\section{Validation with synthetic STIX visibilities}

The standard model for solar flares \citep{1988psf..book.....T,piana2022hard} implies that the emitting configurations may involve a coronal loop corresponding to thermal energies, which, in several cases, is characterized by an elliptic shape; and two circular sources at higher, non-thermal energies (depending on the position of the foot-points with respect to the limb, just one source may be visible to the telescope). Accounting for this scenario, we have generated synthetic sets of STIX visibilities corresponding to five different configurations:
\begin{enumerate}
    \item A single circular Gaussian source.
    \item A single Gaussian elliptic source.
    \item A single loop, obtained by bending the elliptic source.
    \item Two approaching circular Gaussian foot-points.
    \item Two circular Gaussian sources with significantly different size.
    \item A configuration made of one big loop and a small circular Gaussian source.
\end{enumerate}
In order to quantitatively assess the performances of CLEAN and multi-scale CLEAN we have considered the capability of the algorithms to reconstruct the following imaging parameters:
\begin{itemize}
    \item The total flux of the emitting source.
    \item The position of the flux peak.
    \item The values of the Full Width at Half Maximum (FWHM) along both axes.
    \item The maximum intensity of each source.
\end{itemize}
After giving specific values to these parameters, we have generated the ground-truth maps using the Particle Swarm Optimization algorithm \citep{2022A&A...668A.145V}, and providing each map with $10,000$ counts s$^{-1}$ cm$^{-2}$ arcsec$^{-2}$ keV$^{-1}$. Then, we have computed the corresponding STIX synthetic visibility sets by computing the Fourier transform of these images sampled in correspondence of the $(u,v)$ points characterizing the STIX design. Specifically, we have assumed that just collimators from $3$ through $10$ were available, since the calibration process concerned with the highest resolution collimators is not yet finalized. 

\begin{figure*}
    \centering   \includegraphics[width=0.80\linewidth]{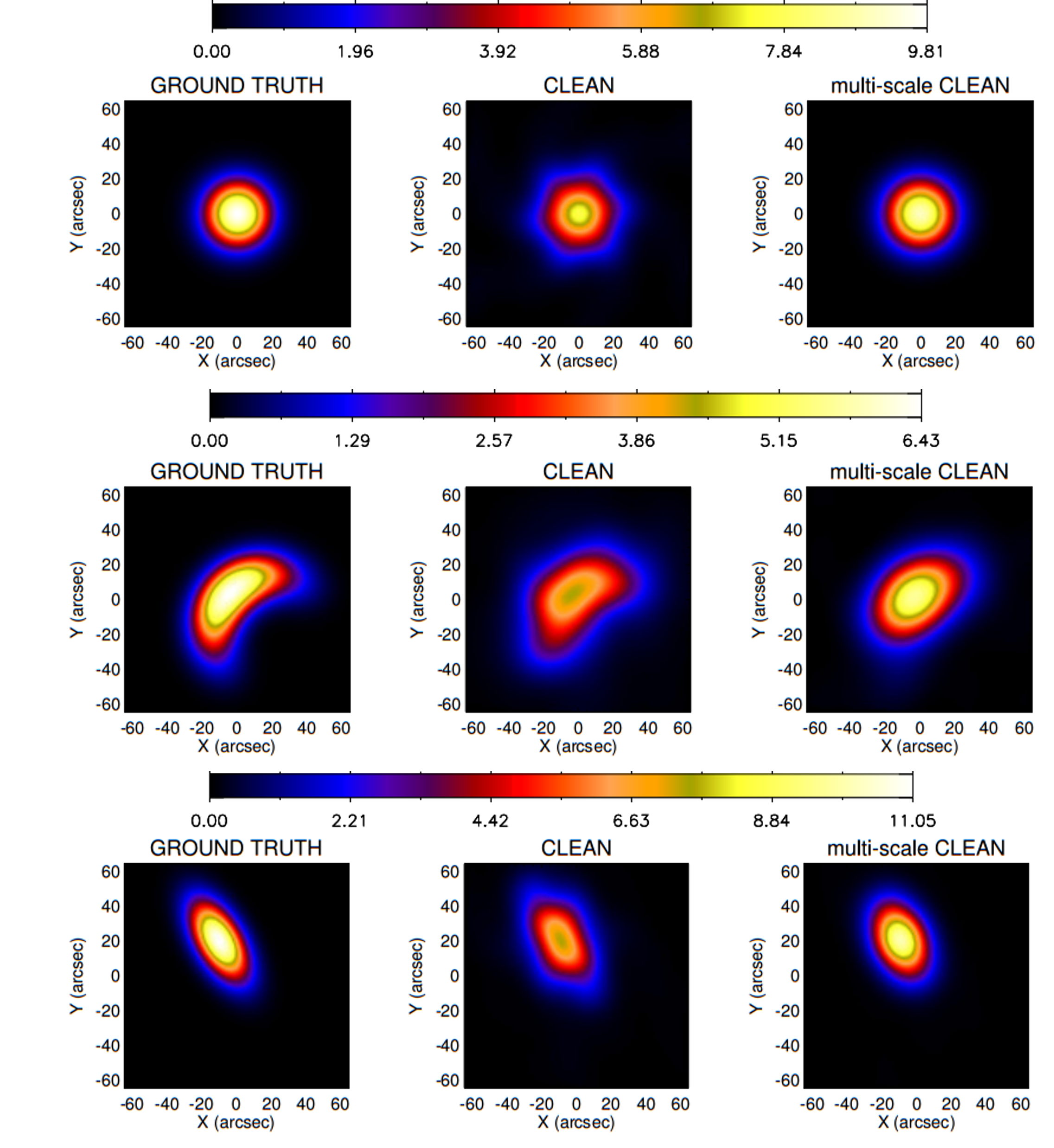}
    \caption{Reconstruction of a single source: comparison between the ground truth (first column) and the CLEAN and multi-scale CLEAN performances (second and third column, respectively). First row: a circular Gaussian source; second row: a loop-wise source; third row: an elliptical source. The corresponding imaging parameters are in Table \ref{tab:single}. }
    \label{fig:single}
\end{figure*}

Figure \ref{fig:single} and Table \ref{tab:single} qualitatively and quantitatively compare the reconstructions provided by CLEAN and multi-scale CLEAN with the ground-truth, in the case of the three single-source configurations. As far as multi-scale CLEAN is concerned, in all cases we have chosen a two-scale setup. Specifically, in the cases of the circular Gaussian and loop-like sources the two scales have been realized by grouping detectors 3 through 9 for the high resolution scale, and using detector 10 for the low resolution one; in the case of the elliptical source, the two scales have been obtained by grouping detectors 3 through 7 and detectors 8 through 10, respectively. These results show that both CLEAN and its multi-scale version are able to model the correct shape of the flaring sources, although multi-scale CLEAN performs better than standard CLEAN in the estimate of the total and maximum fluxes. Interestingly, these results of multi-scale CLEAN have been obtained by setting the algorithm in a two-scale configuration, which shows its flexibility in adapting the setting performances to the input data.

\begin{figure*}
    \centering   \includegraphics[width=0.80\linewidth]{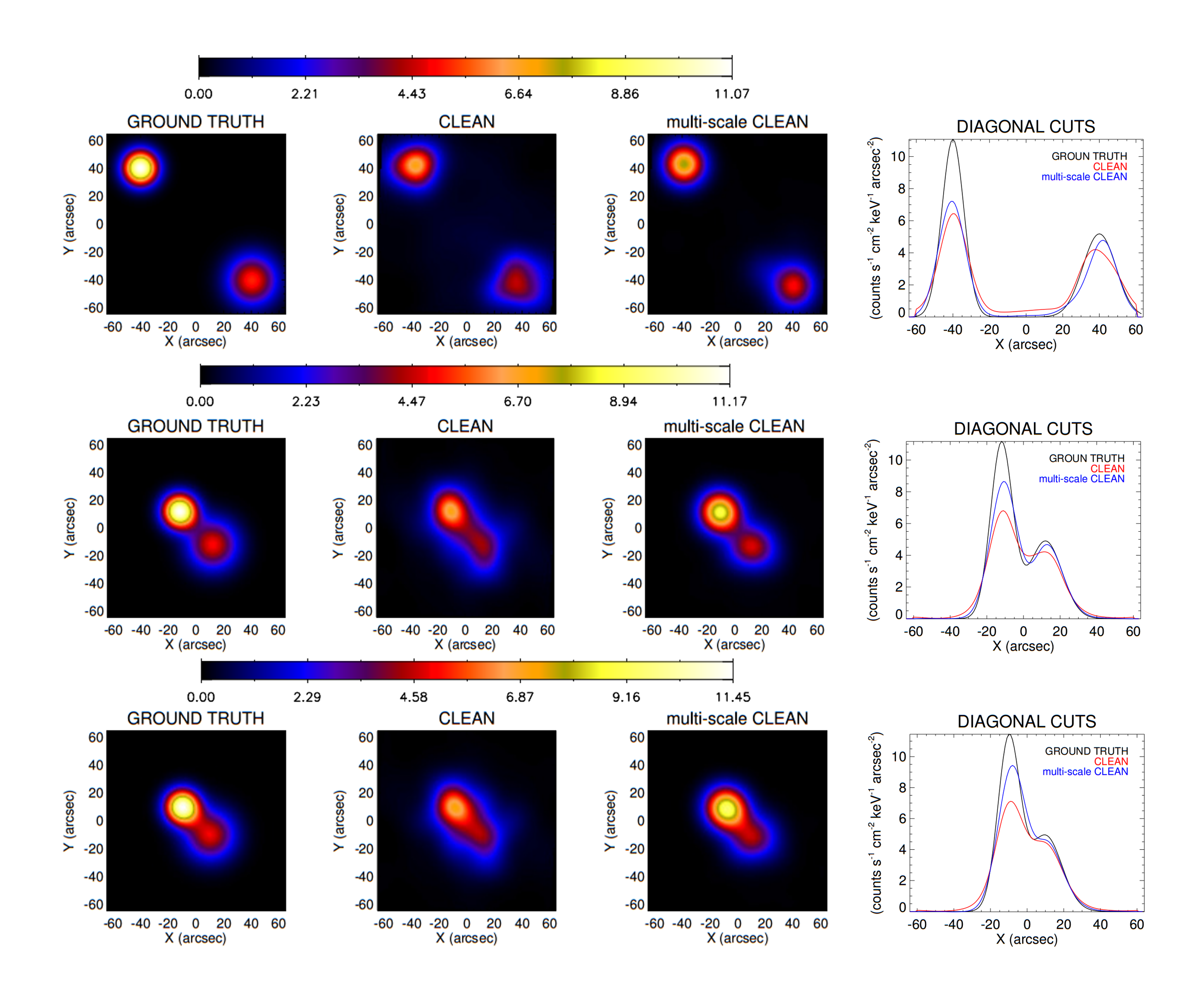}
    \caption{Reconstruction of two approaching circular Gaussian sources: comparison between the ground truth (first column) and the CLEAN and multi-scale CLEAN performances (second and third column, respectively). The fourth column compares the profiles along the cut across the source peaks.}
    \label{fig:double}
\end{figure*}

We have then considered three two-source configurations, in which two Gaussian sources with equal total flux are gradually brought closer and closer. Once again, the total flux to reconstruct is $10^4$ counts s$^{-1}$ cm$^{-2}$ arcsec$^{-2}$ keV$^{-1}$ and multi-scale CLEAN has been applied in a two-scale setup (detectors 3 through 7 for the first scale, and detectors 8 through 10 for the second scale). The results in Figure \ref{fig:double} and Table \ref{tab:double} show that both CLEAN and multi-scale CLEAN are able to reconstruct the morphology of the two-source configurations, although the multi-scale algorithm better preserves the circular shapes. Further, the parametric performances of multi-scale CLEAN are significantly more reliable, with this algorithm being more effective in reproducing both the total flux and the local maxima. In particular, the values of the Full Width at Half Maximum (FWHM) provided by multi-scale CLEAN better recover the ones associated with the ground truth, showing that the multi-scale release improves the under-resolution issue characterizing standard CLEAN performances.

\begin{figure*}
    \centering   \includegraphics[width=0.80\linewidth]{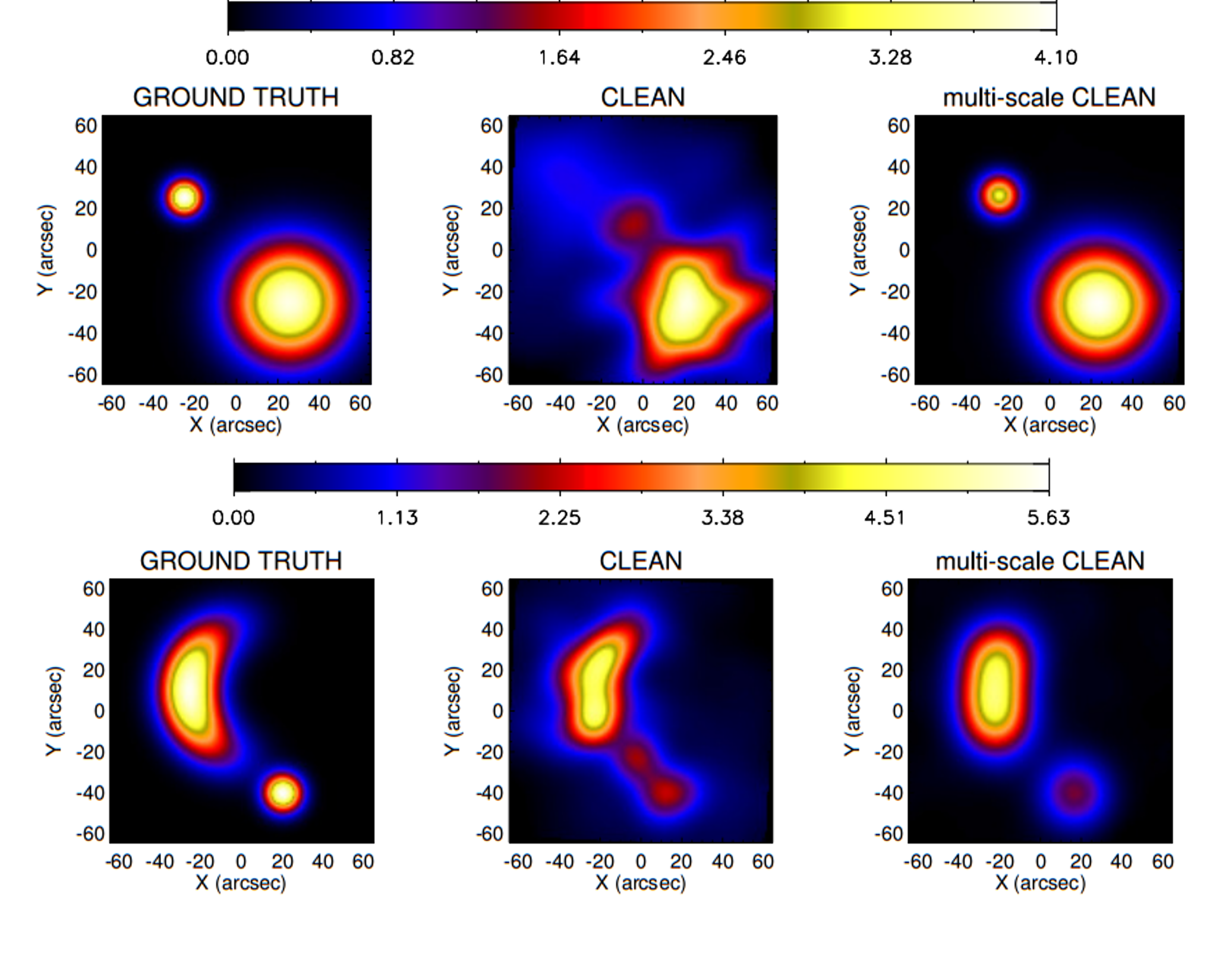}
    \caption{Reconstruction of two sources with different sizes: comparison between the ground truth (first column) and the CLEAN and multi-scale CLEAN reconstructions (second and third column, respectively). In the first row both sources are modelled by a circular Gaussian function, while in the second row the biggest source has a loop shape.}
    \label{fig:double2}
\end{figure*}

The final analysis in the synthetic setting was concerned with two sources with significantly different sizes and same peak intensity, which is probably the most indicative test to validate the effectiveness of a multiscale approach. In the case of the reconstruction of the two circular-source configuration, multi-scale CLEAN has been applied using three scales, obtained by grouping detectors 3 and 4 (first scale), detectors 5 through 8 (second scale), and detectors 9 and 10 (third scale). For the reconstruction of the configuration made of the big loop and the small circular source, we have used two scales made of detectors 3 through 8 and detectors 9 and 10, respectively. Table \ref{tab:double2} shows the input and reconstructed parameters corresponding to the reconstructions in Figure \ref{fig:double2}. Once again, the multi-scale approach outperforms standard CLEAN as far as the ability to estimate the imaging parameters is concerned, although it has some difficulties in accurately reproduce the exact loop shape. 

\begin{table*}
\centering\footnotesize
\begin{tabular}{c c c c c c c}
\hline\hline
circle  & total flux & max & $x_{max}$ & $y_{max}$ & FWHM$_x$ & FWHM$_y$  \\
  \hline
ground truth & $10000$ & $9.81$ & $-1$ & $-1$ & $30$ & $30$\\
CLEAN & 11353 & $7.66$ & $-1$ & $-1$ & $33$ & $33$ \\
multi-scale CLEAN & $10167$ & $9.04$ & $-1$ & $-1$ & $31$ & $31$ \\
\hline
ellipse &  &&&&& \\
\hline
 ground truth & $10000$ & $11.05$ & $-11$ & $19$ & $20$ & $40$ \\
 CLEAN & $10978$ & $7.49$ & $-11$ & $19$ & $37$ & $27$ \\
multi-scale CLEAN & $10103$ & $10.06$ & $-11$ & $19$ & $31$ & $25$ \\
\hline
loop &&&&&& \\
\hline
ground truth & $10000$ & $6.43$ & $-5$ & $3$ & $50$ & $30$ \\
CLEAN & $11209$ & $4.36$ & $-5$ & $2$ & $43$ & $49$ \\
multi-scale CLEAN & $10079$ & $5.65$ & $-4$ & $0$ & $37$ & $39$ \\
\hline
\end{tabular}
\caption{The imaging parameters characterizing the ground truth and the CLEAN and multi-scale CLEAN reconstructions in the case of the three single source configurations (see Figure \ref{fig:single}).}\label{tab:single}
\end{table*}

\begin{table*}
\centering\tiny
\resizebox{17.5cm}{!}{
\begin{tabular}{c c c c c c c c c c c c c }
\hline\hline
& & \multicolumn{5}{ c }{First source} & & \multicolumn{5}{ c }{Second source} 
\\
\cline{3-7}
\cline{9-13}
  & total flux & max & $x_{max}$ & $y_{max}$ & FWHM$_x$ & FWHM$_y$ & &max & $x_{max}$ & $y_{max}$ & FWHM$_x$ & FWHM$_y$ \\
  \hline 
ground truth & $10000$ & $11.07$ & $-41$ & $39$ & $20$ & $20$ & & $5.18 $ & $39$ & $-41$ & $30$ & $30$ \\
CLEAN & $13172$ & $6.86$ & $ -40 $ & $39$ & $24$ & $27$ & & $4.45$ & $41$ & $-43$ & $33$ & $35$ \\
multi-scale CLEAN & $9813$ & $7.83$ & $-41$ & $40$ & $24$ & $24$ & & $5.10$ & $41$ & $-43$ & $27$ & $26$ \\
\hline
ground truth & $10000$ & $11.17$ & $-12$ & $11$ & $20$ & $20$ & & $4.91$ & $11$ & $-12$ & $30$ & $30$ \\
CLEAN & $11693$ & $6.89$ & $-11$ & $10$ & $28$ & $25$ & & $4.27$ & $12$ & $-12$ & $37$ & $19$ \\
multi-scale CLEAN & $10085$ & $8.78$ & $-11$ & $10$ & $24$ & $24$ & & $4.75$ &  $12$ & $-12$ & $27$ & $21$ \\
\hline
ground truth & $10000$ & $11.45$ & $-10$ & $9$ & $20$ & $20$ & & $4.95$ & $9$ & $-10$ & $30$ & $30$ \\
CLEAN & $11456$ & $7.16$ & $-9$ & $8$ & $29$ & $28$ & & $4.56$ & $7$ & $-7$ & $40$ & $17$ \\
multi-scale CLEAN & $9936$ & $9.79$ & $-8$ & $7$ & $25$ & $25$  & &$4.71$ & $7$ & $-7$ & $40$ & $17$ \\
\hline
\end{tabular}}
\caption{The imaging parameters characterizing the ground truth and the CLEAN and multi-scale CLEAN reconstructions in the case of two approaching circular Gaussian sources (see Figure \ref{fig:double}). }\label{tab:double}
\end{table*}

\begin{table*}
\centering\tiny
\resizebox{17.5cm}{!}{
\begin{tabular}{c c c c c c c c c c c c c }
\hline\hline
& & \multicolumn{5}{ c }{First source} & & \multicolumn{5}{ c }{Second source} 
\\
\cline{3-7}
\cline{9-13}
  & total flux & max & $x_{max}$ & $y_{max}$ & FWHM$_x$ & FWHM$_y$ & &max & $x_{max}$ & $y_{max}$ & FWHM$_x$ & FWHM$_y$ \\
  \hline 
ground truth & $10000$ & $4.10$ & $-26$ & $24$ & $14$ & $14$ & & $3.96$ & $24$ & $-26$ & $46$ & $46$ \\
CLEAN & $13884$ & $1.53$ & $-6$ & $11$ & $28$ & $40$ & & $3.96$ & $20$ & $-26$ & $47$ & $50$ \\
multi-scale CLEAN & $9785$ & $3.09$ & $-25$ & $25$ & $17$ & $17$ & & $4.06$ &  $22$ & $-27$ & $45$ & $45$ \\
\hline
ground truth & $10000$ & $5.51$ & $-27$ & $9$ & $50$ & $30$ & & $5.63$ & $19$ & $-41$ & $15$ & $15$ \\
CLEAN & $13156$ & $4.75$ & $-24$ & $-1$ & $50$ & $24$ & & $2.26$ & $11$ & $-41$ & $28$ & $27$ \\
multi-scale CLEAN & $10021$ & $4.73$ & $-23$ & $7$ & $52$ & $29$ & &$1.91$ & $16$ & $-41$ & $29$ & $28$ \\
\hline
\end{tabular}}
\caption{The imaging parameters characterizing the ground truth and the CLEAN and multi-scale CLEAN reconstructions in the case of two sources with different sizes (see Figure \ref{fig:double2}).}\label{tab:double2}
\end{table*}

\section{Analysis of real flaring events}

The application of CLEAN and multi-scale CLEAN to visibilities recorded by STIX is concerned with three events occurred on February, 24 2024, March, 10 2024, and May, 14 2024, respectively. All three events were observed from both Earth orbits and the Solar Orbiter. Therefore, for each event, we generated the EUV map from data recorded by SDO/AIA at 1600 $\AA$ (for the February 24-2024 and March 10-2024 flares) and at 131 $\AA$ (for the May 14-2024 flare). Further, we considered measurements of the flaring radiation collected by the Hard X-ray Imager (HXI) on-board the Chinese cluster ASO-S, which leverages an imaging modality very similar to the one utilized by the ESA telescope. 

\begin{figure*}
    \centering   \includegraphics[width=0.90\linewidth]{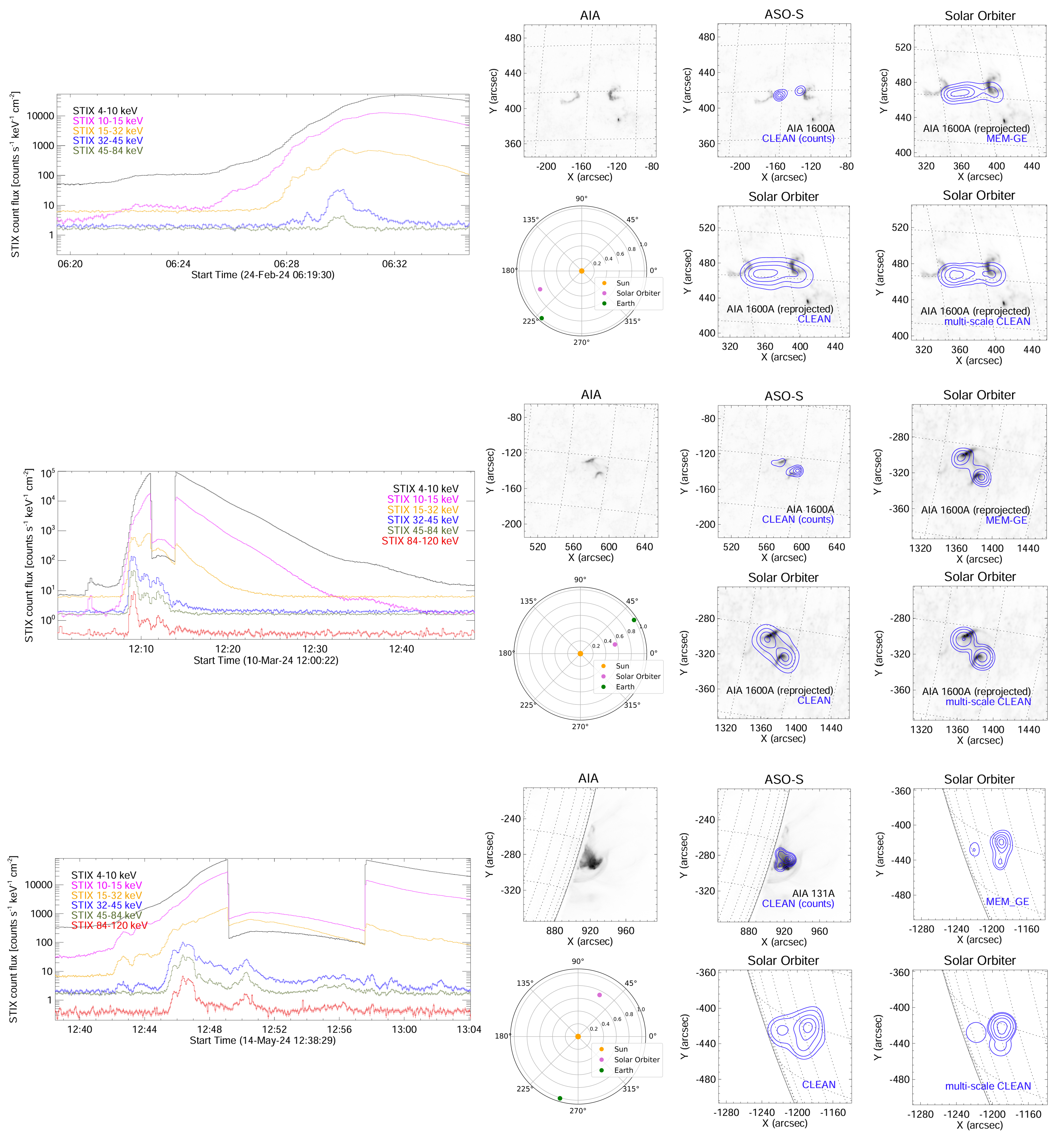}
    \caption{Multi-scale CLEAN in the case of three real events observed by STIX: the February 24 2024 event in the time range 06:27:00 - 06:29:00 UT and for the 25 - 50 keV energy channel (first row); the March 10 2024 event in the time range 12:05:00 - 12:07:00 UT and for the 22 - 50 keV energy channel (second row); the May 14 2024 event in the time range 12:46:55 - 12:47:55 UT for the 32 - 76 keV (third row). The first column contains the light-curves. The second column provides: the AIA map at 1600 $\AA$ in the case of February 24 and March 10, at 131 $\AA$ in the case of May 14 (top left panel); the HXI level curves reconstructed by CLEAN (top middle panel); the STIX level curves reconstructed by MEM$\_$GE (top right panel); the position of the Earth and of Solar Orbiter at the event (bottom left panel); the STIX level curves reconstructed by CLEAN (bottom middle panel); the STIX level curves reconstructed by multi-scale CLEAN (bottom right panel). All level curves are superimposed onto the AIA map. The HXI observations are made in the 06:31:02 - 06:31:42 UT time interval and for the 25 - 50 keV energy channel for the first event; in the 12:10:00 - 12:11:00 UT time interval and for the 22 - 50 keV energy channel for the second event; in the 12:46:45 - 12:47:00 UT time interval and for the 35 - 55 keV energy channel for the third event.}
    \label{fig:fig-5}
\end{figure*}

The results of this analysis are shown in Figure \ref{fig:fig-5}  which allows for a qualitative assessment of CLEAN and multi-scale CLEAN performances for the interpretation of STIX experimental visibilities. In fact, for each data set, we have superimposed the level curves corresponding to the CLEAN and multi-scale CLEAN reconstructions onto the AIA EUV maps that have been appropriately reprojected in order to account for the different relative positions of the two instruments with respect to the flaring source. The same procedure was followed by using the level curves provided by 
\begin{itemize}
\item A constrained maximum entropy method implemented in the MEM$\_$GE routine \citep{2020ApJ...894...46M}, which is often considered as the imaging benchmark by the STIX community. 
\item A CLEAN algorithm applied on a data set recorded by HXI in (approximately) the same time range and for the same energy channel.
\end{itemize}
The CLEAN method used for the analysis of the HXI observations used counts as input, since the calibration process for the Chinese instrument is not yet finalized. The quantitative comparison of the different imaging approaches was performed in Table \ref{tab:parameters-alt} by computing the $\chi^2$ values predicted by the CLEAN component map and the CLEAN, multi-scale CLEAN and MEM$\_$GE reconstructions in the case of STIX visibilities. For HXI we did not compute any statistical metric, because the count uncertainties are not yet available for that instrument, and the visibility uncertainties were approximated as 10$\%$ of the maximum observed visibility amplitude. Since these error estimates are not derived from the actual noise properties of the data, the resulting metrics (both $\chi^2$ and C-statistic) cannot be interpreted as statistically meaningful confidence levels.

\begin{table}[t]
\centering
\scriptsize
\begin{tabular}{ccc}
Event & Method & $\chi^2$ \\
\hline
SOL2024-02-24T06 & & \\
    & CLEAN Components    & 0.95 \\
    & CLEAN Map           & 4.74 \\
    & multi-scale CLEAN   & 1.02 \\
    & MEM\_GE             & 3.09 \\
\hline
SOL2024-03-10T12 & & \\
    & CLEAN Components    & 1.43 \\
    & CLEAN Map           & 21.77 \\
    & multi-scale CLEAN   & 2.64 \\
    & MEM\_GE             & 6.20 \\
\hline
SOL2024-05-14T12 & & \\
    & CLEAN Components    & 2.14 \\
    & CLEAN Map           & 4.70 \\
    & multi-scale CLEAN   & 3.28 \\
    & MEM\_GE             & 2.78 \\
\hline
\end{tabular}
\caption{Reduced $\chi^2$ values obtained with different imaging methods for the three flaring events in Figure \ref{fig:fig-5}. The first column reports the event date, the second the imaging method, and the third the corresponding reduced $\chi^2$ value.}
\label{tab:parameters-alt}
\end{table}




\section{Conclusions}
This study introduces a computational tool for image reconstruction from STIX observations that is based on a multi-scale version of CLEAN, and compares its performances with the ones provided by standard CLEAN deconvolution. The analysis of synthetic visibilities shows that multi-scale CLEAN outperforms standard CLEAN in the ability to qualitatively reproduce the ground-truth shape even in the case of single-scale flaring sources. However, from a more quantitative point of view, the parameters that describe the different properties of the source are reconstructed with similar precision. In the case of experimental measurements, the reconstructions provided by multi-scale CLEAN are clearly better resolved than the ones provided by CLEAN and rather similar to the ones corresponding to the use of MEM$\_$GE. This trend is quantitatively confirmed by the values of the corresponding goodness of fit measured by $\chi^2$ (in the case of CLEAN, multi-scale CLEAN, and MEM$\_$GE applied to STIX visibilities).

The next step for the development of the multi-scale approach for STIX visibilities will be the integration of the tool\footnote{Multi-scale CLEAN codes are available at the MIDA group GitHub \url{https://github.com/theMIDAgroup/MultiScaleCLEAN_codes}).} with an algorithm able to automatically choose the number of scales based on the input data.

\section*{Acknowledgements}
Solar Orbiter is a space mission of international collaboration between ESA and NASA, operated by ESA. The STIX instrument is an international collaboration between Switzerland, Poland, France, Czech Republic, Germany, Austria, Ireland, and Italy. AV, MP, and AMM are supported by the 'Accordo ASI/INAF Solar Orbiter: Supporto scientifico per la realizzazione degli strumenti Metis, SWA/DPU e STIX nelle Fasi D-E'. MP acknowledges the support of the PRIN PNRR 2022 Project 'Inverse Problems in the Imaging Sciences (IPIS)' 2022ANC8HL, cup: D53D23005740006. The research by AV, AMM, and MP was performed within the framework of the MIUR Excellence Department Project awarded to Dipartimento di Matematica, Università di Genova, CUP D33C23001110001. PM is supported by the Swiss National Science Foundation in the framework of the project Robust Density Models for High Energy Physics and Solar Physics (rodem.ch), CRSII5\_193716. AV, AMM, and MP also acknowledge the support of the Fondazione Compagnia di San Paolo within the framework of the Artificial Intelligence Call for Proposals, AIxtreme project (IDRol: 71708).  AV also aknowledges the GNCS-INdAM PIANIS project “Problemi Inversi e Approssimazione Numerica in Imaging Solare”.

\bibliographystyle{aa}
\bibliography{sample631}

\begin{thebibliography}{25}
\expandafter\ifx\csname natexlab\endcsname\relax\def\natexlab#1{#1}\fi

\bibitem[{{Aschwanden} {et~al.}(2002){Aschwanden}, {Schmahl}, \&
  {Team}}]{2002SoPh..210..193A}
{Aschwanden}, M.~J., {Schmahl}, E., \& {Team}, T.~R. 2002, Solar Physics, 210,
  193

\bibitem[{{Benvenuto} {et~al.}(2013){Benvenuto}, {Schwartz}, {Piana}, \&
  {Massone}}]{2013A&A...555A..61B}
{Benvenuto}, F., {Schwartz}, R., {Piana}, M., \& {Massone}, A.~M. 2013,
  Astronomy and Astrophysics, 555, A61

\bibitem[{{Benz} {et~al.}(2012){Benz}, {Krucker}, {Hurford}, {Arnold},
  {Orleanski}, {Gr{\"o}belbauer}, {Klober}, {Iseli}, {Wiehl}, {Csillaghy},
  {Etesi}, {Hochmuth}, {Battaglia}, {Bednarzik}, {Resanovic}, {Grimm},
  {Viertel}, {Commichau}, {Meuris}, {Limousin}, {Brun}, {Vilmer}, {Skup},
  {Graczyk}, {Stolarski}, {Michalska}, {Nowosielski}, {Cichocki}, {Mosdorf},
  {Seweryn}, {Przepi{\'o}rka}, {Sylwester}, {Kowalinski}, {Mrozek},
  {Podgorski}, {Mann}, {Aurass}, {Popow}, {Onel}, {Dionies}, {Bauer},
  {Rendtel}, {Warmuth}, {Woche}, {Pl{\"u}schke}, {Bittner}, {Paschke},
  {Wolker}, {Van Beek}, {Farnik}, {Kasparova}, {Veronig}, {Kienreich},
  {Gallagher}, {Bloomfield}, {Piana}, {Massone}, {Dennis}, {Schwarz}, \&
  {Lin}}]{2012SPIE.8443E..3LB}
{Benz}, A.~O., {Krucker}, S., {Hurford}, G.~J., {et~al.} 2012, in Society of
  Photo-Optical Instrumentation Engineers (SPIE) Conference Series, Vol. 8443,
  Space Telescopes and Instrumentation 2012: Ultraviolet to Gamma Ray, ed.
  T.~{Takahashi}, S.~S. {Murray}, \& J.-W.~A. {den Herder}, 84433L

\bibitem[{Bertero(2006)}]{bertero2006regularization}
Bertero, M. 2006, in Inverse Problems: Lectures given at the 1st 1986 Session
  of the Centro Internazionale Matematico Estivo (CIME) held at Montecatini
  Terme, Italy, May 28--June 5, 1986 (Springer), 52--112

\bibitem[{{Bong} {et~al.}(2006){Bong}, {Lee}, {Gary}, \&
  {Yun}}]{2006ApJ...636.1159B}
{Bong}, S.-C., {Lee}, J., {Gary}, D.~E., \& {Yun}, H.~S. 2006, Astrophysical
  Journal, 636, 1159

\bibitem[{{Dennis} \& {Pernak}(2009)}]{2009ApJ...698.2131D}
{Dennis}, B.~R. \& {Pernak}, R.~L. 2009, Astrophysical Journal, 698, 2131

\bibitem[{{Duval-Poo} {et~al.}(2018){Duval-Poo}, {Piana}, \&
  {Massone}}]{2018A&A...615A..59D}
{Duval-Poo}, M.~A., {Piana}, M., \& {Massone}, A.~M. 2018, Astronomy and
  Astrophysics, 615, A59

\bibitem[{{Felix} {et~al.}(2017){Felix}, {Bolzern}, \&
  {Battaglia}}]{2017ApJ...849...10F}
{Felix}, S., {Bolzern}, R., \& {Battaglia}, M. 2017, Astrophysical Journal,
  849, 10

\bibitem[{{Hannah} {et~al.}(2008){Hannah}, {Christe}, {Krucker}, {Hurford},
  {Hudson}, \& {Lin}}]{2008ApJ...677..704H}
{Hannah}, I.~G., {Christe}, S., {Krucker}, S., {et~al.} 2008, Astrophysical
  Journal, 677, 704

\bibitem[{{H{\"o}gbom}(1974)}]{1974A&AS...15..417H}
{H{\"o}gbom}, J.~A. 1974, Astronomy and Astrophysics Suppl. Ser., 15, 417

\bibitem[{{Hurford} {et~al.}(2002){Hurford}, {Schmahl}, {Schwartz}, {Conway},
  {Aschwanden}, {Csillaghy}, {Dennis}, {Johns-Krull}, {Krucker}, {Lin},
  {McTiernan}, {Metcalf}, {Sato}, \& {Smith}}]{2002SoPh..210...61H}
{Hurford}, G.~J., {Schmahl}, E.~J., {Schwartz}, R.~A., {et~al.} 2002, Solar
  Physics, 210, 61

\bibitem[{Kosugi {et~al.}(1992)Kosugi, Sakao, Masuda, Makishima, Inda,
  Murakami, Ogawara, Yaji, \& Matsushita}]{kosugi1992hard}
Kosugi, T., Sakao, T., Masuda, S., {et~al.} 1992, PASJ: Publications of the
  Astronomical Society of Japan (ISSN 0004-6264), vol. 44, no. 5, p. L45-L49.,
  44, L45

\bibitem[{{Krucker} {et~al.}(2020){Krucker}, {Hurford}, {Grimm}, {K{\"o}gl},
  {Gr{\"o}belbauer}, {Etesi}, {Casadei}, {Csillaghy}, {Benz}, {Arnold},
  {Molendini}, {Orleanski}, {Schori}, {Xiao}, {Kuhar}, {Hochmuth}, {Felix},
  {Schramka}, {Marcin}, {Kobler}, {Iseli}, {Dreier}, {Wiehl}, {Kleint},
  {Battaglia}, {Lastufka}, {Sathiapal}, {Lapadula}, {Bednarzik}, {Birrer},
  {Stutz}, {Wild}, {Marone}, {Skup}, {Cichocki}, {Ber}, {Rutkowski}, {Bujwan},
  {Juchnikowski}, {Winkler}, {Darmetko}, {Michalska}, {Seweryn}, {Bia{\l}ek},
  {Osica}, {Sylwester}, {Kowalinski}, {{\'S}cis{\l}owski}, {Siarkowski},
  {St{\k{e}}{\'s}licki}, {Mrozek}, {Podg{\'o}rski}, {Meuris}, {Limousin},
  {Gevin}, {Le Mer}, {Brun}, {Strugarek}, {Vilmer}, {Musset}, {Maksimovi{\'c}},
  {F{\'a}rn{\'\i}k}, {Koz{\'a}{\v{c}}ek}, {Ka{\v{s}}parov{\'a}}, {Mann},
  {{\"O}nel}, {Warmuth}, {Rendtel}, {Anderson}, {Bauer}, {Dionies}, {Paschke},
  {Pl{\"u}schke}, {Woche}, {Schuller}, {Veronig}, {Dickson}, {Gallagher},
  {Maloney}, {Bloomfield}, {Piana}, {Massone}, {Benvenuto}, {Massa},
  {Schwartz}, {Dennis}, {van Beek}, {Rodr{\'\i}guez-Pacheco}, \&
  {Lin}}]{2020A&A...642A..15K}
{Krucker}, S., {Hurford}, G.~J., {Grimm}, O., {et~al.} 2020, Astronomy and
  Astrophysics, 642, A15

\bibitem[{{Lemen} {et~al.}(2012){Lemen}, {Title}, {Akin}, {Boerner}, {Chou},
  {Drake}, {Duncan}, {Edwards}, {Friedlaender}, {Heyman}, {Hurlburt}, {Katz},
  {Kushner}, {Levay}, {Lindgren}, {Mathur}, {McFeaters}, {Mitchell}, {Rehse},
  {Schrijver}, {Springer}, {Stern}, {Tarbell}, {Wuelser}, {Wolfson}, {Yanari},
  {Bookbinder}, {Cheimets}, {Caldwell}, {Deluca}, {Gates}, {Golub}, {Park},
  {Podgorski}, {Bush}, {Scherrer}, {Gummin}, {Smith}, {Auker}, {Jerram},
  {Pool}, {Soufli}, {Windt}, {Beardsley}, {Clapp}, {Lang}, \&
  {Waltham}}]{2012SoPh..275...17L}
{Lemen}, J.~R., {Title}, A.~M., {Akin}, D.~J., {et~al.} 2012, Solar Physics,
  275, 17

\bibitem[{{Lin} {et~al.}(2004){Lin}, {Dennis}, {Hurford}, {Smith}, \&
  {Zehnder}}]{2004SPIE.5171...38L}
{Lin}, R.~P., {Dennis}, B., {Hurford}, G., {Smith}, D.~M., \& {Zehnder}, A.
  2004, in Society of Photo-Optical Instrumentation Engineers (SPIE) Conference
  Series, Vol. 5171, Telescopes and Instrumentation for Solar Astrophysics, ed.
  S.~{Fineschi} \& M.~A. {Gummin}, 38--52

\bibitem[{{Massa} {et~al.}(2020){Massa}, {Schwartz}, {Tolbert}, {Massone},
  {Dennis}, {Piana}, \& {Benvenuto}}]{2020ApJ...894...46M}
{Massa}, P., {Schwartz}, R., {Tolbert}, A.~K., {et~al.} 2020, Astrophysical
  Journal, 894, 46

\bibitem[{{Massone} {et~al.}(2009){Massone}, {Emslie}, {Hurford}, {Prato},
  {Kontar}, \& {Piana}}]{2009ApJ...703.2004M}
{Massone}, A.~M., {Emslie}, A.~G., {Hurford}, G.~J., {et~al.} 2009,
  Astrophysical Journal, 703, 2004

\bibitem[{{Metcalf} {et~al.}(1996){Metcalf}, {Hudson}, {Kosugi}, {Puetter}, \&
  {Pina}}]{1996ApJ...466..585M}
{Metcalf}, T.~R., {Hudson}, H.~S., {Kosugi}, T., {Puetter}, R.~C., \& {Pina},
  R.~K. 1996, Astrophysical Journal, 466, 585

\bibitem[{{Perracchione} {et~al.}(2023){Perracchione}, {Camattari}, {Volpara},
  {Massa}, {Massone}, \& {Piana}}]{2023ApJS..268...68P}
{Perracchione}, E., {Camattari}, F., {Volpara}, A., {et~al.} 2023,
  Astrophysical Journal Suppl. Ser., 268, 68

\bibitem[{Piana {et~al.}(2022)Piana, Emslie, Massone, \&
  Dennis}]{piana2022hard}
Piana, M., Emslie, A.~G., Massone, A.~M., \& Dennis, B.~R. 2022, Hard X-ray
  Imaging of Solar Flares, Vol. 164 (Springer)

\bibitem[{{Schmahl} {et~al.}(2007){Schmahl}, {Pernak}, {Hurford}, {Lee}, \&
  {Bong}}]{2007SoPh..240..241S}
{Schmahl}, E.~J., {Pernak}, R.~L., {Hurford}, G.~J., {Lee}, J., \& {Bong}, S.
  2007, Solar Physics, 240, 241

\bibitem[{{Tandberg-Hanssen} \& {Emslie}(1988)}]{1988psf..book.....T}
{Tandberg-Hanssen}, E. \& {Emslie}, A.~G. 1988, {The physics of solar flares}

\bibitem[{{Volpara} {et~al.}(2024){Volpara}, {Catalano}, {Piana}, \& {Maria
  Massone}}]{2024InvPr..40l5017V}
{Volpara}, A., {Catalano}, M., {Piana}, M., \& {Maria Massone}, A. 2024,
  Inverse Problems, 40, 125017

\bibitem[{{Volpara} {et~al.}(2022){Volpara}, {Massa}, {Perracchione},
  {Francesco Battaglia}, {Garbarino}, {Benvenuto}, {Krucker}, {Piana}, \&
  {Massone}}]{2022A&A...668A.145V}
{Volpara}, A., {Massa}, P., {Perracchione}, E., {et~al.} 2022, Astronomy and
  Astrophysics, 668, A145

\bibitem[{{Zhang} {et~al.}(2019){Zhang}, {Chen}, {Wu}, {Chang}, {Hu}, {Su},
  {Zhang}, {Wang}, {Liang}, {Ma}, {Guo}, {Cai}, {Zhang}, {Huang}, {Peng},
  {Tang}, {Zhao}, {Zhou}, {Wang}, {Song}, {Ma}, {Xu}, {Yang}, {Lu}, {He},
  {Tao}, {Ma}, {Lv}, {Bai}, {Cao}, {Huang}, \& {Gan}}]{2019RAA....19..160Z}
{Zhang}, Z., {Chen}, D.-Y., {Wu}, J., {et~al.} 2019, Research in Astronomy and
  Astrophysics, 19, 160

\end{thebibliography}

\end{document}